\documentclass[prb,eqsecnum,showpacs,twocolumn,amsmath,amssymb]{revtex4}
\usepackage{graphicx}
\usepackage{dcolumn}
\usepackage{bm}
\begin{document}
\title{Instability of the Mott or Lieb-Wu insulator caused by an infinitesimal perturbation
}
\author{Fusayoshi J. Ohkawa}
\affiliation{Division of Physics, Faculty of Science, Hokkaido University, Sapporo 060-0810, Japan}
\email{fohkawa@phys.sci.hokudai.ac.jp}
%
\received{%
28 September 2011
}
%
\begin{abstract} 
The half-filled ground state of the Hubbard model in one dimension is studied by Kondo-lattice theory.
Because of the Kondo effect, any insulating ground state with a complete gap open is unstable in the presence of an infinitesimal perturbation. 
This fact casts doubt on the claim by E. H. Lieb and F. Y. Wu, Phys. Rev. Lett. {\bf 20}, 1445 (1968) that the half-filled ground state is the Mott insulator. 
Though the claim is based on a rigorous result given by the Bethe-ansatz solution, the rigorous result is simply a necessary condition for the ground state being an insulator. 
\end{abstract} 
\pacs{71.10.Fd, 71.30.+h, 75.10.Kt}
\maketitle
 %
\maketitle
\section{Introduction}
\label{SecIntroduction}
Mott localization is still a crucial and present-day issue.\cite{mott}
Whether magnetism is itinerant-electron or local-moment magnetism is directly related to physics of Mott localization.
High-temperature (high-$T_c$) superconductivity occurs in the vicinity of a phase of a local-moment type of antiferromagnetic (AF) insulator.\cite{bednortz}
Then it is anticipated that the mechanism of high-$T_c$ superconductivity is closely related to physics of Mott localization.\cite{Anderson-SC,FJO-SC1,FJO-SC2}

The Hubbard model is one of the simplest effective Hamiltonians to study Mott localization.
According to Hubbard's theory,\cite{Hubbard1,Hubbard2}
when the on-site repulsion $U$ is large enough such that $U\gtrsim 2z|t|$, where $t$ is the transfer integral between nearest neighbors and $z$ is their coordination number, the band splits into two subbands; the subbands are called the upper and lower Hubbard bands and a gap between them is called the Hubbard gap.
According to Gutzwiller's theory,\cite{gutzwiller1,gutzwiller2,gutzwiller3} when the on-site $U$ is large enough and electrons are almost half filled such that $U\gtrsim 2z|t|$ and $N\simeq L$, where $N$ and $L$ are the numbers of electrons and unit cells, respectively, electrons are renormalized into heavy quasi-particles. A mid-band appears at the chemical potential between the upper and lower Hubbard bands;\cite{mid-band} 
the mid-band is called the Gutzwiller band in this paper.
According to Brinkman and Rice's theory,\cite{brinkman}
when $N=L$ 
a metal-insulator (MI) transition occurs at $U_{c}\simeq 2z|t|$.
In a metallic phase at $U< U_{c}$ and $U\simeq U_{c}$, the density of states (DOS) is of a three-peak structure, the Gutzwiller band between the upper and lower Hubbard bands.\cite{mid-band} 
In an insulating phase at $U> U_{c}$, the Gutzwiller band disappears and the Hubbard gap is a complete gap. 
According to these theories, the half-filled GS can be a paramagnetic type of Mott insulator, which is simply called a Mott insulator in this paper, only if $U$ is large enough such that $U>U_c$.
However, any non-half-filled GS is a metal for any $U$.



The GS can also be an insulator if an AF gap opens.\cite{slater} 
In a high-$T$ phase at $T\gg E_{\rm F}^*/k_{\rm B}$, where $E_{\rm F}^*$ is the effective Fermi energy and $k_{\rm B}$ is the Boltzmann constant, electrons behave like local moments even if no gap opens.
If $U\gtrsim 2z|t|$ and $E_{\rm F}^* \ll z|t|$, the high-$T$ phase is in a sense the Mott insulator. 
Then a local-moment type of AF insulator, in which the N\'{e}el temperature $T_{\rm N}$ is as high as $T_{\rm N} \gg E_{\rm F}^*/k_{\rm B}$, is an AF type of Mott insulator.
An itinerant-electron type of AF insulator, in which $T_{\rm N} \ll E_{\rm F}^*/k_{\rm B}$, is not any type of Mott insulator. 


In spite of a lot of studies,
the nature of the Mott insulator at $T=0\hskip2pt$K
is not clear in the least except for the opening of the Hubbard gap.
For example, it is controversial whether the MI transition at $U_c$ is a first-order or second-order one, or a discontinuous or continuous one.
According to Brinkman and Rice's theory,\cite{brinkman} either of the specific-heat coefficient and the spin susceptibility diverges as $U\rightarrow U_{c}-0$.
This suggests that the transition is continuous. However, no symmetry seems to be broken in the Mott insulator at $U\ge U_c$; residual entropy seems to remain and the third law of thermodynamics
seems to be broken in it. This suggests that the transition is discontinuous. On the other hand, it is unlikely that the third law is broken when $U$ is finite. Since the appearance of rigidity due to symmetry breaking can make the third law valid in an ordered state,\cite{AndersonText} it is possible that a {\it hidden} order exists in the Mott insulator.
It is controversial whether or not residual entropy remains in the Mott insulator.
Under the present circumstances, 
it is desirable to critically examine whether the GS can be the Mott insulator, from the very beginning.

When $U$ is infinite and $N=L$ in the canonical ensemble, 
each unit cell is occupied by a single electron with an arbitrary spin and there is no empty nor double occupancy. 
No electron is itinerant; 
no bosonic charge excitation is possible or the charge gap is infinitely large, while all the bosonic spin excitations are degenerate at the zero energy.
This is a prototypic Mott insulator.
Since its residual entropy is $k_{\rm B}\ln 2$ per unit cell,
even the GS of it is in a sense a high-$T$ phase.
When electrons are removed from it, {\it empty occupancies} or {\it holes} are itinerant.\cite{comInfU-1D} 
Since electrons become more itinerant as $U$ becomes smaller, any non-half filled GS must be a metal for any finite or infinite $U$, at least if no symmetry is broken.

The Hubbard model in one dimension is particular: No symmetry can be broken in it,\cite{mermin} and the Bethe-ansatz solution for it was given by Lieb and Wu.\cite{lieb-wu}
Denote the GS energy by $E_{\rm G}(N)$ as a function of $N$. 
Two {\it chemical potentials} are defined for the GS of $N$ electrons: 
\begin{subequations}
\begin{align}
\mu_{+}(N)= E_{\rm G}(N+1)- E_{\rm G}(N),
\end{align}
for the addition of an electron and 
\begin{align}
\mu_{-}(N)= E_{\rm G}(N)- E_{\rm G}(N-1),
\end{align}
\end{subequations}
for the removal of an electron. 
Then a {\em gap} is defined by 
\begin{align}
\epsilon_{\rm g} (N) =\mu_{+}(N)-\mu_{-}(N).
\end{align}
When $N \ne L$, $\epsilon_{\rm g}(N)=0$. 
Then, any non-half-filled GS is a metal,\cite{comMS} i.e., a Tomonaga-Luttinger (TL) electron liquid.\cite{Tomonaga,Luttinger,Mattis}
When $N=L$, on the other hand,
\begin{align}\label{EqCond2}
\epsilon_{\rm g}(L)>0,
\end{align}
for $U/|t|>0$.
Based on this rigorous result, Lieb and Wu claimed that an MI transition occurs at $U_c/|t|=0$ and the half-filled GS is the Mott insulator for any nonzero $U$; this insulator is also called a Lieb-Wu insulator in this paper.


When $U$ is finite, the third law of thermodynamics is not broken
in the Bethe-ansatz solution.
It sounds curious that
the nature of the Mott or Lieb-Wu insulator is different from that of the prototypic Mott insulator, in which the third law is broken. 
The rigorous result (\ref{EqCond2}) is no sufficient condition for the GS being an insulator but simply a necessary condition for it; e.g., $\epsilon_{\rm g}(L)$ is nonzero in a metallic fine particle because of the long-range Coulomb interaction. 
It is desirable to reexamine whether the half-filled GS in one dimension is really an insulator, in particular, 
in the grand canonical ensemble; electrodes are necessary to measure the static conductivity of a system and the electrodes play a role of an electron reservoir for the system.

Not only $E_{\rm G}(L)$ but also $\epsilon_{\rm g}(L)$ as a function of $U$ are singular at $U=0$;\cite{lieb-wu,takahashi}
they cannot be expanded in terms of $U$.
Thus, $\epsilon_{\rm g}(L)$ is extremely or singularly small for $U/|t|\ll 1$:
\begin{align}\label{EqSmallEg}
\lim_{U/|t|\rightarrow 0} \epsilon_{\rm g}(L)/(U/|t|)^\eta = 0,
\end{align}
for any real $\eta\ge 0$,
though it is as large as the gap given by Hubbard's theory, or the Hubbard gap, for $U\gtrsim 2|t|$.
Since no symmetry is broken, no rigidity can appear in the gapped phase characterized by $\epsilon_{\rm g}(L)>0$.
Since it is anticipated that such a non-rigid gapped phase is unstable at least in the limit of $U/|t|\rightarrow 0$ in the presence of a perturbation such as one due to an electron reservoir, 
it is interesting to examine how large $U_c$ can be as a function of the strength of the perturbation.

In the grand canonical ensemble,
the averaged electron number $\langle N \rangle$ can be an irrational one even in the limit of $T\rightarrow0\hskip1pt$K; thus, it is a continuous function of the chemical potential.
This means that quantum fluctuations of $N$ are {\it effectively} considered in the conventional formulation, though an electron reservoir is only {\it implicitly} considered in it.
Quantum fluctuations of $N$ are effective at least when $\left<N\right>$ is a non-integer, e.g., even when $\left<N\right>=L\pm 0$.
However, they may or may not be effective when exactly $\left<N\right>=L$:
If they are not effective,
the half-filled GS is composed of only components of $N=L$ electrons, so that 
it is anticipated that the gapped phase is rigid and $U_c/|t|=0$, i.e., the half-filled GS is an insulator for any nonzero $U$ in the grand canonical ensemble; though the GS is a metal even for $\left<N\right>=L\pm 0$, it is an insulator for exactly $\left<N\right>=L$.\cite{comBandInsulator}
On the other hand,
if quantum fluctuations of $N$ are still effective even when exactly $\left<N\right>=L$, the half-filled GS is composed of not only components of $N=L$ electrons but also those of $N\ne L$ electrons.
Since any non-half-filled GS of $N\ne L$ electrons is a metal in the canonical ensemble, it is anticipated that the gapped phase is not rigid and $U_c/|t|=+\infty$, i.e., the half-filled GS is a metal for any finite $U$ in the presence of an electron reservoir or in the grand canonical ensemble.
It is a crucial issue which is the truth.

Based on Kondo-lattice theory (KLT),\cite{Mapping-1,Mapping-2,Mapping-3}
it was shown in a previous paper\cite{toyama} that the GS cannot be 
the Mott insulator for any filling $\langle N \rangle$ in the presence of a simplified electron reservoir, though it is never any actual type of reservoir or electrodes.
In this theory,\cite{toyama} the Kondo effect plays a crucial role in causing any insulating GS that is possible within the constrained Hilbert subspace where no symmetry is allowed to be broken to be unstable.
The main purpose of this paper is to show that because of the Kondo effect the half-filled GS in one dimension cannot be the Mott or Lieb-Wu insulator in the presence of an infinitesimal perturbation, 
though it is a phenomenological one that guarantees the average $\left<N\right>$ to be able to be an irrational number. 
Another purpose is to examine or discuss whether the half-filled GS is the Mott or Lieb-Wu insulator in the absence of the infinitesimal perturbation.

This paper is organized as follows:
Preliminary is given in Sec.\hskip2pt\ref{SecPreliminaries}.
Result is given in Sec.\hskip2pt\ref{SecResult}.
Discussion is given in Sec.\hskip2pt\ref{SecDiscussion}.
Conclusion is given in Sec.\hskip2pt\ref{SecConclusion}.
In Appendix\hskip2pt\ref{SecRVB}, it is shown that 
the resonating-valence-bond (RVB) mechanism\cite{fazekas} stabilizes a metallic GS rather than the Mott insulator.
In Appendix\hskip2pt\ref{SecContinousModel}, it is shown that the Kondo effect is irrelevant in a non-lattice model or a continuous model.
In Appendix\hskip2pt\ref{SecInfDim}, it is shown that
the Mott insulator in infinite dimensions is also unstable in the presence of the infinitesimal perturbation.

\section{Preliminary}
\label{SecPreliminaries}
\subsection{Kondo-lattice theory}
\label{SecKL-theory}
We consider the Hubbard model with a finite onsite repulsion $U$ in one dimension:
\begin{equation}\label{EqHubbardH}
{\cal H} = 
- t \sum_{\left<ij\right>\sigma} 
d_{i\sigma}^\dag d_{j\sigma}
+U \sum_i n_{i\uparrow}n_{i\downarrow},
\end{equation}
where $n_{i\sigma}=d_{i\sigma}^\dag d_{i\sigma}$ 
and $\left<ij\right>$ stands for nearest neighbors; the other notations are conventional. 
An infinitesimal perturbation is phenomenologically considered. 

Since no symmetry can be broken in one dimension, \cite{mermin} the Hilbert space is constrained within the subspace where no symmetry is allowed to be broken. 
Then the thermal Green function of electrons is given by
\begin{align}\label{EqGreen}
G_{\sigma}(i\varepsilon_l, k) =
\frac1{\displaystyle
i\varepsilon_l \hskip-1pt + \hskip-1pt \mu \hskip-1pt 
- \hskip-1pt E(k) \hskip-1pt - \hskip-1pt
\Sigma_{\sigma}(i\varepsilon_l, k) \hskip-1pt + \hskip-1pt
i\delta \frac{\varepsilon_l}{|\varepsilon_l|} },
\end{align}
where $\varepsilon_l=(2l+1)\pi k_{\rm B}T$, with $l$ being an integer, is a fermionic energy, $\mu$ is the chemical potential,
\begin{align}\label{EqDispD=1}
E(k) = -2t \cos(k a), 
\end{align}
with $a$ being the lattice constant, and
$\Sigma_\sigma(i\varepsilon_l, k)$ is the self-energy.
The term of $i\delta\hskip1pt\varepsilon_l/|\varepsilon_l|$ is due to the phenomenological perturbation; the limit of $\delta\rightarrow 0$ or $\delta=0^+$ is assumed in this paper. 
The main purpose of this paper is to show that the half-filled GS cannot be an insulator in the presence of $\delta=0^+$.

In general, the self-energy is decomposed into the single-site $\tilde{\Sigma}_\sigma(i\varepsilon_l)$ and the multisite $\Delta\Sigma_\sigma(i\varepsilon_l, k)$:
\begin{align}
\Sigma_\sigma(i\varepsilon_l, k) =
\tilde{\Sigma}_\sigma(i\varepsilon_l)
+ \Delta\Sigma_\sigma(i\varepsilon_l, k).
\end{align}
According to the Kondo-lattice theory (KLT)\cite{Mapping-1,Mapping-2,Mapping-3,toyama} or 
dynamical mean-field theory (DMFT),\cite{georges}
calculating the single-site $\tilde{\Sigma}_\sigma(i\varepsilon_l)$ is mapped to a problem of self-consistently determining the Anderson model (AM) and solving it, i.e., a self-consistent Kondo problem.

We consider the AM defined by
\begin{align}\label{EqAnderson}
\tilde{\cal H} &=
\tilde{\epsilon}_d \sum_{\sigma}\tilde{n}_{d\sigma}
+ \sum_{{\bf k}\sigma} \tilde{E}_c({\bf k}) \tilde{c}_{{\bf k}\sigma}^\dag \tilde{c}_{{\bf k}\sigma}
+ \tilde{U} \tilde{n}_{d\uparrow} \tilde{n}_{d\downarrow}
\nonumber \\ & \quad
+ \frac1{\sqrt{ \tilde{L} }} \sum_{{\bf k}\sigma} \left(
\tilde{V}_{\bf k}\tilde{c}_{{\bf k}\sigma}^\dag \tilde{d}_\sigma
+ \tilde{V}_{\bf k}^* \tilde{d}_\sigma^\dag \tilde{c}_{{\bf k}\sigma}\right) ,
\end{align} 
where 
$\tilde{n}_{d\sigma}=\tilde{d}_{\sigma}^\dag \tilde{d}_{\sigma}$; the notations here are also conventional except for the use of tildes. 
The parameters of the AM are determined in a way such that the self-energy of it is equal to the single-site $\tilde{\Sigma}_\sigma(i\varepsilon_l)$ of the Hubbard model. Then the Green function of the AM is given by
\begin{align}\label{EqGreenAnderson}
\tilde{G}_{\sigma}(i\varepsilon_l) &=
\frac1{\displaystyle
i\varepsilon_l \hskip-1pt + \hskip-1pt
\tilde{\mu} \hskip-1pt - \hskip-1pt \tilde{\epsilon}_d 
\hskip-1pt - \hskip-1pt
\tilde{\Sigma}_\sigma(i\varepsilon_l)
\hskip-1pt - \hskip-1pt \frac1{\pi} \hskip-2pt \int_{-\infty}^{+\infty}\hskip-12pt d \epsilon \hskip2pt
\frac{\tilde{\Delta}(\epsilon)}
{i\varepsilon_l \hskip-1pt - \hskip-1pt \epsilon }
}, 
\end{align}
where $\tilde{\mu}$ is the chemical potential and 
\begin{equation}\label{EqDelta}
\tilde{\Delta}(\varepsilon) =
\frac{\pi}{\tilde{L}} \sum_{\bf k} |\tilde{V}_{\bf k}|^2 
\delta\bigl[\varepsilon+\tilde{\mu}- \tilde{E}_c({\bf k})\bigr] .
\end{equation}
The Fermi surface (FS) is defined by
$\tilde{\mu}=\tilde{E}_c({\bf k})$.
If the FS exists, then 
$\tilde{\Delta}(0)>0$ 
unless $|\tilde{V}_{\bf k}|^2$ vanishes on the whole of the FS.

For the Hubbard model, a single-site Feynman diagram is defined as one such that it is only composed of on-site 
interaction and site-diagonal electron lines in the site representation: $U$ and 
\begin{align}\label{EqR}
R_{\sigma}(i\varepsilon_l) = 
\frac1{L}\sum_{k}G_{\sigma}(i\varepsilon_l,k).
\end{align}
For the AM, there is a Feynman diagram corresponds to the single-site one. It is composed of on-site interaction and electron lines: 
$\tilde{U}$ and $\tilde{G}_{\sigma}(i\varepsilon_l)$.
The condition to determine the AM is so simple that
$U=\tilde{U}$ 
and 
\begin{align}\label{EqMap2}
R_{\sigma}(i\varepsilon_l)=
\tilde{G}_{\sigma}(i\varepsilon_l).
\end{align}
The condition (\ref{EqMap2}) is equivalent to a set of 
$\mu= \tilde{\mu}- \tilde{\epsilon}_d$
and
\begin{align}\label{EqMap4}
\tilde{\Delta}(\varepsilon) &=
{\rm Im}\left[
\tilde{\Sigma}_\sigma(\varepsilon+i0)
+1/R_{\sigma}(\varepsilon+i0) \right] .
\end{align} 
where $\tilde{\Sigma}_\sigma(\varepsilon+i0)$ and $R_{\sigma}(\varepsilon+i0)$ are the analytical continuations of $\tilde{\Sigma}_\sigma(i\varepsilon_l )$ and $R_{\sigma}(i\varepsilon_l)$, respectively, from the upper half plane onto the real axis.
Since $U=\tilde{U}$ and $\mu= \tilde{\mu}- \tilde{\epsilon}_d$ are trivial,
Eq.\hskip1pt(\ref{EqMap4}) is a practical mapping condition to determine the AM. 

If the multisite $\Delta\Sigma_\sigma(i\varepsilon_l, k)$ is ignored in this theoretical framework, the theory is simply within the supreme single-site approximation (S$^3$A),\cite{Metzner,Muller-H1,Muller-H2,Janis}  
which is rigorous in infinite dimensions within the constrained Hilbert subspace where no symmetry is allowed to be broken.
Either the DMFT or dynamical coherent potential approximation (DCPA) \cite{dcpa} is also within the S$^3$A.
In the KLT of this paper, a set of $\tilde{\Delta}(\varepsilon)$, $\tilde{\Sigma}_\sigma(\varepsilon+i0)$, and $\Delta\Sigma_\sigma(\varepsilon+i0,k)$ should be self-consistently considered and determined to satisfy Eq.\hskip1pt(\ref{EqMap4}).

Though the contribution of a Feynman diagram for the self-energy may diverge as $T\rightarrow0\hskip2pt$K and $|\varepsilon|\rightarrow0$ or $|\varepsilon_l|\rightarrow0$ for finite $l$, 
it is finite for $T>0\hskip2pt$K because no symmetry is broken in one dimension. 
Therefore, it is certain that the mapping is possible at least for $T>0\hskip2pt$K.
When the summation of all the Feynman diagrams is followed by taking the limit of $T\rightarrow 0\hskip2pt$K, 
the mapping is possible even in the limit of $T\rightarrow 0\hskip2pt$K.


Because of Eq.\hskip1pt(\ref{EqMap2}),
the DOS of the Hubbard model is equal to that of the mapped AM:
\begin{subequations}\label{EqDOS}
\begin{align}\label{EqDOS1}
\rho(\varepsilon) &= 
- \frac1{\pi L}\sum_{k} {\rm Im}
\hskip1pt G_\sigma(\varepsilon+i0,k)
\\ \label{EqDOS2} &= 
-\frac1{\pi} {\rm Im}
\tilde{G}_{\sigma}(\varepsilon+i0).
\end{align}
\end{subequations}
If $\rho(0)>0$, the GS is a metal. If $\rho(0)=0$, the GS is a non-metal: an insulator or a zero-gap semiconductor.

The averaged electron number is given by
\begin{align}
\left<{\cal N}\right> &=
\int_{-\infty}^{+\infty} \hskip-8pt d\varepsilon f(\varepsilon)\rho(\varepsilon),
\end{align}
where
\begin{align}
{\cal N}=\sum_{i\sigma}n_{i\sigma},
\end{align}
$\langle\cdots\rangle$ stands for the statistical average, and
\begin{align}\label{EqFD-F}
f(\varepsilon) =
\frac1{e^{\varepsilon/(k_{\rm B}T)}+1}.
\end{align}
In the thermodynamic limit of $L\rightarrow+\infty$ followed by the limit of $\delta\rightarrow0$, 
$\left<{\cal N}\right>$ can be an irrational number and is a continuous function of $\mu$ even in the limit of $T\rightarrow0\hskip2pt$K.

When $\mu = U/2$, both the Hubbard model and the mapped AM are half-filled and symmetrical:
\begin{align}\label{EqMuEdU}
\mu = \tilde{\mu} - \tilde{\epsilon}_d = U/2 =\tilde{U}/2,
\end{align}
%
$\langle {\cal N}\rangle/L =\langle \tilde{n}_{\uparrow}+ \tilde{n}_{\downarrow}\rangle=1$,
%
$\rho(\varepsilon)=\rho(-\varepsilon)$,
%
\begin{align}\label{EqDeltaSym}
\tilde{\Delta}(\varepsilon)=\tilde{\Delta}(-\varepsilon) ,
\end{align}
and so on.
It follows from Eq.(\ref{EqDeltaSym}) that 
\begin{align}\label{EqSymDelta}
- \frac1{\pi} \hskip-2pt \int_{-\infty}^{+\infty} \hskip-12pt d \epsilon \hskip2pt
\frac{\tilde{\Delta}(\epsilon)}
{+i0 - \epsilon } = i \tilde{\Delta}(0).
\end{align}
In a self-consistent solution, $\tilde{\Sigma}_\sigma(\varepsilon+i0)$ or $\Sigma_\sigma(\varepsilon+i0,k)$ may be continuous or discontinuous at $\varepsilon=0$. In either case,
\begin{align}\label{EqNatureSA}
\tilde{\mu}-\tilde{\epsilon}_d -{\rm Re}\tilde{\Sigma}_\sigma(+i0)=0,
\end{align}
and
\begin{align}\label{EqNatureSH}
\mu -E(\pm k_{\rm F})-{\rm Re}\Sigma_\sigma(+i0,\pm k_{\rm F})=0,
\end{align}
where $k_{\rm F}=\pi/2a$, so that $E(\pm k_{\rm F})=0$. 

In this paper, the half-filled GS in the symmetrical Hubbard model is mainly studied; 
$0\le U/|t|<+\infty$, $L\rightarrow+\infty$, and $T\rightarrow 0\hskip1pt$K are assumed.

\subsection{Kondo effect}
\label{SecKondo}
The Kondo problem for the $s$-$d$ model and the AM has already been solved:\cite{yosida,poorman,wilsonKG,nozieres,yamada1,yamada2,exact1,exact2,exact3,exact4}
The single energy scale of $k_{\rm B}T_{\rm K}$ only appears in low-energy phenomena; $T_{\rm K}$ is called the Kondo temperature.
In this paper, it is defined by
\begin{align}\label{EqDefTK}
k_{\rm B}T_{\rm K}=
1/\left[\tilde{\chi}_s(0)\right]_{T\rightarrow0\hskip1pt{\rm K}},
\end{align}
where $\tilde{\chi}_s(0)$ is the static susceptibility of localized electrons in the AM; in this paper, the susceptibility is defined in a way such that it does not include the conventional factor $g^2\mu_{\rm B}^2/4$, where $g$ and $\mu_{\rm B}$ are the $g$ factor and the Bohr magneton, respectively.
The Kondo temperature depends on $\tilde{\Delta}(\varepsilon)$ or $\tilde{\Delta}(0)$, which is defined by Eq.\hskip2pt(\ref{EqDelta}):
If the FS exists and 
\begin{align}\label{EqPositiveDelta}
\tilde{\Delta}(0)>0,
\end{align}
then $T_{\rm K}>0\hskip1pt$K. 
If $\tilde{\Delta}(0)=0$, $T_{\rm K}>0\hskip1pt$K or $T_{\rm K}=0\hskip1pt$K.
The nature of the GS depends on $T_{\rm K}$:
If $T_{\rm K}>0\hskip1pt$K, the GS is a singlet and a normal Fermi liquid (FL). If $T_{\rm K}=0\hskip1pt$K, the GS is neither a singlet nor a normal FL; residual entropy remains 
and localized electrons behave like a local moment at any temperature $T$.

According to the FL theory by Yamada and Yosida\cite{yamada1,yamada2} and the Bethe-ansatz solution \cite{exact1,exact2,exact3,exact4} for the AM, the adiabatic continuation as a function of $\tilde{U}$ holds in the region of $0\le \tilde{U}/[\pi\tilde{\Delta}(0)]<+\infty$; i.e., the transition between the weak-coupling region of $\tilde{U}/[\pi\tilde{\Delta}(0)]\ll 1$ and the strong-coupling region of $\tilde{U}/[\pi\tilde{\Delta}(0)]\gg 1$ is only a crossover and there is no finite critical $\tilde{U}$.
In this paper, therefore, physics in the AM for any $\tilde{U}$ is simply called the Kondo effect.

In the self-consistent Kondo problem of this paper,
eventual $\tilde{\Delta}(\varepsilon)$ and $T_{\rm K}$ are 
renormalized by intersite effects, so that they should be 
self-consistently determined with the multisite $\Delta\Sigma_\sigma(\varepsilon+i0, k)$ to satisfy Eq.\hskip1pt(\ref{EqMap4}). 
The nature of the GS of the mapped AM depends on the eventual and self-consistent $\tilde{\Delta}(0)$ and $T_{\rm K}$.

\section{Result}
\label{SecResult}
\subsection{Case of $\delta=0^+$}
\label{SecResult1}
According to Eq.\hskip1pt(\ref{EqMap4}), it follows that 
\begin{align}\label{EqProofDelta1}
\tilde{\Delta}(\varepsilon) 
\ge \delta,
\end{align}
for any $\varepsilon$;
the proof of Eq.\hskip1pt(\ref{EqProofDelta1}) is in parallel with that in Appendix of the previous paper.\cite{toyama}
It should be noted that this inequality holds even if the self-energy $\tilde{\Sigma}_\sigma(\varepsilon+i0)$ or $\Delta\Sigma_\sigma(\varepsilon + i0,k)$ is anomalous.
If $\delta>0$, then $\tilde{\Delta}(\varepsilon)>0$; thus, $T_{\rm K}>0\hskip2pt$K for any self-consistent solution, though $T_{\rm K}$ may be infinitesimally low for $\delta=0^+$.

When $\delta=0^+$, the GS of the mapped AM is a normal FL because of the Kondo effect. 
In the presence of an infinitesimal magnetic field $H$ 
in the AM,
$\tilde{\Sigma}_\sigma(\varepsilon+i0)$ of the normal FL in the AM can be expanded in terms of $\varepsilon/(k_{\rm B}T_{\rm K})$ in a normal way such that \cite{yamada1,yamada2}
\begin{align}\label{EqExpansion}
\tilde{\Sigma}_\sigma(\varepsilon + i0) &=
\frac1{2}U + \bigl(1 - \tilde{\phi}_\gamma\bigr) \varepsilon
+ \bigl(1-\tilde{\phi}_s\bigr) \frac1{2}\sigma g \mu_{\rm B} H
\nonumber \\ & \quad 
- i \tilde{\phi}_{\tau} \varepsilon^2/(k_{\rm B}T_{\rm K})
+ O\left(\varepsilon^3\right),
\end{align}
%
where $\tilde{\phi}_\gamma$, $\tilde{\phi}_s$,  and $\tilde{\phi}_{\tau}$ are positive constants. The Wilson ratio is defined by
\begin{align}\label{EqWilsonRatio}
\tilde{W}_s= \tilde{\phi}_s/\tilde{\phi}_\gamma,
\end{align}
which is used in Appendix\hskip2pt\ref{SecRVB}. 
In this section, $H=0$ is assumed. It follows that
\begin{align}\label{EqG1}
G_\sigma(\varepsilon+i0,k) &=
1\big/ \bigl[\tilde{\phi}_\gamma \varepsilon -E(k) 
+ i \tilde{\phi}_{\tau} \varepsilon^2/(k_{\rm B}T_{\rm K})
\nonumber \\ & \quad
- \Delta\Sigma_\sigma(\varepsilon+i0, k) 
+ O\bigl(\varepsilon^3\bigr) \bigr].
\end{align}

In one dimension, the charge-spin separation occurs in pair or bosonic excitations\cite{Tomonaga,Luttinger,Mattis} and a single-particle fermionic excitation is in a sense a composite of charge and spin excitations;\cite{luther,mattis}
however, it is doubtful or at least unclear whether charge-spin separation occurs in the fermionic single-particle excitation.
In any case, it is inevitable that the multi-site $\Delta\Sigma_\sigma(\varepsilon+i0, k)$ is anomalous at $\varepsilon=0$ in one dimension.

If $\Delta\Sigma_\sigma(\varepsilon+i0, k)$ is continuous and finite at $\varepsilon=0$, it follows from Eqs.\hskip1pt(\ref{EqDOS1}), (\ref{EqNatureSH}), and (\ref{EqG1}) that
\begin{subequations}\label{EqRho>0}
\begin{align}\label{EqRho>0-a}
\lim_{\varepsilon\rightarrow 0} \rho(\varepsilon)>0,
\end{align}
and
\begin{align}\label{EqRho>0-b}
\rho(0)>0, 
\end{align}
\end{subequations}
for the Hubbard model.
It also follows from Eqs.\hskip1pt(\ref{EqGreenAnderson}), (\ref{EqDOS2}), (\ref{EqSymDelta}), (\ref{EqNatureSA}), and (\ref{EqRho>0}) that
\begin{align}\label{EqRhoDelta}
\rho(0) =1/[\pi \tilde{\Delta}(0)]>0,
\end{align}
for the AM.
Since $k_{\rm B}T_{\rm K}$ is the only low-energy scale in the AM, Eq.(\ref{EqRhoDelta}) implies that 
$\rho(\varepsilon)>0$ for at least $|\varepsilon|\lesssim k_{\rm B}T_{\rm K}$.
Thus, the GS is a metal in this case.

Because of the Kramers-Kronig relation, if one of the real and imaginary parts of $\Delta\Sigma_\sigma(\varepsilon+i0, k)$ is discontinuous at $\varepsilon=0$ the other is divergent at $\varepsilon=0$.
%
Even if $\Delta\Sigma_\sigma(\varepsilon+i0, k)$ is divergent at $\varepsilon=0$, no complete gap opens because of the imaginary term of $i \tilde{\phi}_{\tau} \varepsilon^2/(k_{\rm B}T_{\rm K})$ in the single-site self-energy but only a zero-gap can open.
For example, if $\Delta\Sigma_\sigma(\varepsilon+i0, k)$ has a pole at $\varepsilon=-i0$ in a way such that
\begin{align}
\Delta\Sigma_\sigma(\varepsilon+i0, k) = 
c_1/(\varepsilon +i0) + \cdots, 
\end{align}
then
\begin{align}
\rho(\varepsilon) =
\bigl[\tilde{\phi}_{\tau}/(c_1^2k_{\rm B}T_{\rm K})\bigr] \varepsilon^4 + O\left(\varepsilon^6\right),
\end{align}
i.e., only a zero-gap opens.
Even if $\Delta\Sigma_\sigma(\varepsilon+i0, k)$ is divergent at $\varepsilon=0$, the GS cannot be an insulator with a complete gap open but it can be only a zero-gap semiconductor.

When $U$ is finite and $\delta=0^+$, the half-filled GS is a metal or a zero-gap semiconductor and it can never be any insulator with a complete gap open, so that $U_c/|t|=+\infty$. 
The only approximation used to obtain this conclusion is that the expansion (\ref{EqExpansion}) is made use of. 
If the expansion coefficients of $\tilde{\phi}_\gamma$, $\tilde{\phi}_s$, and $\tilde{\phi}_{\tau}$ are rigorous, the expansion becomes rigorous in the limit of $\varepsilon/(k_{\rm B}T_{\rm K})\rightarrow 0$. Essentially, no approximation is used to obtain the conclusion; thus, it is unquestionable and absolute.

\subsection{Case of absolutely $\delta=0$}
\label{SecResult2}
%
%
When $U$ is finite and $\delta=0$, there are two possibilities for $T_{\rm K}$: $T_{\rm K}>0\hskip2pt$K and $T_{\rm K}=0\hskip2pt$K. 
Provided that $T_{\rm K}>0\hskip2pt$K, the expansion (\ref{EqExpansion}) is relevant, so that the single-site self-energy is normal. As proved in Sec.\hskip2pt\ref{SecResult1}, no complete gap opens because of the Kondo effect;
the GS is a metal if the multisite self-energy $\Delta\Sigma_\sigma(\varepsilon+i0, k)$ is continuous and finite at $\varepsilon=0$ or a zero-gap semiconductor if $\Delta\Sigma_\sigma(\varepsilon+i0, k)$ is discontinuous or divergent at $\varepsilon=0$.

Provided that $T_{\rm K}=0\hskip2pt$K, the expansion (\ref{EqExpansion}) is irrelevant, so that not only the multisite self-energy but also the single-site self-energy may be anomalous. Then, the total self-energy $\Sigma_\sigma(\varepsilon+i0, k)$ may be so anomalous that it has a pole just below the real axis in the vicinity of $\varepsilon=0$ and its imaginary part ${\rm Im}\Sigma_\sigma(\varepsilon+i0, k)$ is absolutely zero for at least $- \epsilon_{\rm G}/2 \le \varepsilon\le \epsilon_{\rm G}/2$ except for the existence of the delta function at the singular point where the pole exists; e.g.,
\begin{align}
\Sigma_\sigma(\varepsilon+i0, k) &=
\frac{c_1(k)}{\varepsilon+i0}
+\frac1{\pi}\hskip-1pt\int_{-\infty}^{+\infty} \hskip-10pt d\epsilon
\frac{A(\epsilon;k)}{\varepsilon+i0- \epsilon},
\end{align}
where $A(\epsilon;k)=0$ for at least $- \epsilon_{\rm G}/2 \le \epsilon\le \epsilon_{\rm G}/2$.
A complete gap as large as $\epsilon_{\rm G}$ can only open in such an anomalous case. 
Thus, the GS can be an insulator only if $T_{\rm K}=0\hskip2pt$K; $T_{\rm K}=0\hskip2pt$K is a necessary condition for the GS being an insulator with a complete gap open.\cite{comExoticMetal} 
Essentially, no approximation is used to obtain this conclusion either; thus, it is also unquestionable and absolute.

The only possible insulating GS is one such that it is characterized by $T_{\rm K}=0\hskip2pt$K.
What is proved in Sec.\hskip2pt\ref{SecResult1} is simply that
the insulating GS is unstable if the infinitesimal $\delta=0^+$ is once introduced.

\section{Discussion}
\label{SecDiscussion}
It is proved in Sec.\hskip2pt\ref{SecResult} that the half-filled GS is no insulator in the presence of the phenomenological $\delta=0^+$.
In this section, it is examined or discussed from various points of view whether or not the half-filled GS is an insulator in the absence of $\delta=0^+$.

In field theory based on the grand canonical ensemble, 
the analytical continuation from the upper half-plane onto the real axis is made in a way such that %
\begin{align}
i\varepsilon_l \rightarrow \varepsilon+i\delta_{\rm ac} , 
\end{align}
with $\delta_{\rm ac}= 0^+$.
This $\delta_{\rm ac}=0^+$ plays a similar role to $\delta=0^+$ in this paper. 
If a simplified electron reservoir is considered as in the previous paper,\cite{toyama} the thermal Green function is given instead of Eq.\hskip2pt(\ref{EqGreen}) by 
\begin{align}\label{EqGreenToyama}
G_{\sigma}(i\varepsilon_l, k) =
\frac1{\displaystyle
i\varepsilon_l \hskip-1pt + \hskip-1pt \mu \hskip-1pt 
- \hskip-1pt E(k) \hskip-1pt - \hskip-1pt
\Sigma_{\sigma}(i\varepsilon_l, k) \hskip-1pt + \hskip-1pt
\Gamma(i\varepsilon_l ) },
\end{align}
where $\Gamma(i\varepsilon_l)$ is due to the hybridization with the reservoir. If the FS exists in the reservoir, it follows that
\begin{align}
{\rm Im} \Gamma(+i0) >0.
\end{align}
This ${\rm Im} \Gamma(+i0) >0$ also plays a similar role to $\delta=0^+$ in this paper. 
It is plausible that $T_{\rm K}>0\hskip2pt$K and the third law of thermodynamics is not broken even in the absence of $\delta=0^+$, unless $U$ is infinite. Thus, it is desirable to critically examine whether the half-filled GS is really an insulator with a complete gap open in the absence of the infinitesimal phenomenological $\delta=0^+$

Not only the result of Sec.\hskip2pt\ref{SecResult} but also the argument above in this section cast doubt on whether Eq.\hskip2pt(\ref{EqCond2}) or $\epsilon_{\rm g}(L)>0$ is really a proof that the half-filled GS for $\delta=0$ is an insulator. 
When $T_{\rm K}=0\hskip2pt$K, the residual entropy of localized electrons in the AM is $O(k_{\rm B})$, 
so that the single-site residual entropy of the Hubbard model is also $O(k_{\rm B})$ per unit cell. 
If an insulating GS is possible when $\delta=0$, it should be one such that $T_{\rm K}=0\hskip2pt$K and the third law of thermodynamics is broken in it.
On the other hand, the third law is not broken in the Bethe ansatz solution, which implies that $T_{\rm K}>0\hskip2pt$K even for $\delta=0$. Thus, it is doubtful whether the half-filled GS of the Bethe-ansatz solution is the only possible insulating GS, which is characterized by $T_{\rm K}=0\hskip2pt$K.
The rigorous result of $\epsilon_{\rm g}(L)>0$ is no proof that the half-filled GS of the Bethe-ansatz solution is the Mott insulator.

When an electron or {\it hole} is added to the Hubbard model in the canonical ensemble, the whole of it remains within the Hubbard model.
When it is added to the Hubbard model in the grand canonical ensemble and if quantum fluctuations of ${\cal N}$ are at least effectively considered in a way such that $\left<{\cal N}\right>$ can be an irrational number,
not only a process where the whole or almost whole of it remains within the Hubbard model but also a process is possible such that only an infinitesimally small fraction of it remains within the Hubbard model and the almost whole of it escapes to a reservoir.
The single-particle excitation spectrum $\rho(\varepsilon)$ in the grand canonical ensemble can be different from the spectrum of adding or removing an electron in the canonical ensemble.
Thus, $\epsilon_{\rm g}(L)>0$ is no sufficient condition for 
the opening of a complete gap in $\rho(\varepsilon)$ but
simply a necessary condition for it.

The total lattice of the Hubbard model can be divided into sub-lattices in a way such that a central one is surrounded by the others. 
The others can play a role of an electron reservoir for the central sub-lattice. 
Denote the numbers of electrons and unit cells within the central sub-lattice by $N^\prime$ and $L^\prime$, respectively, and the averaged number of $N^\prime$ by $\langle N^\prime \rangle$: $\langle N^\prime \rangle=L$.
When $U$ is finite, quantum fluctuations of $N^\prime$ develop to a greater or lesser degree and they never vanish.\cite{comBandInsulator} 
The half-filled GS on the central sub-lattice is therefore composed of components of various $N^\prime$ such as $N^\prime=L^\prime$, $L^\prime\pm 1$, $L^\prime\pm 2$, and so on.
Any non-half-filled GS, where $N^\prime\ne L^\prime$, is a metal.
Since the half-filled GS on the central sub-lattice includes such non-half-filled metallic components, it is likely to be a metal.
Since any sub-lattice can be a central one, the half-filled GS on it is also likely to be a metal. 
The half-filled GS on the total lattice is composed of such half-filled metals on the sub-lattices.
Thus, the half-filled GS is likely to be a metal even in the canonical ensemble.

According to a perturbative treatment,\cite{supJ-PWA} when $1 \lesssim U/|t|<+\infty$, the superexchange interaction arises from quantum fluctuations allowing empty and double occupancies.
 The same one can also be derived by a treatment of field theory.\cite{supJ-fjo1,supJ-fjo2} Since it arises from the virtual exchange of a pair excitation of an electron in the upper Hubbard band and a hole in the lower Hubbard band, it works in either a metallic or insulating phase. In the limit of $U/|t|\rightarrow +\infty$, its constant is given by 
\begin{align}\label{EqSuperJ}
J = -4t^2/U,
\end{align}
between nearest neighbors; when $U/|t|\gtrsim 1$, Eq.\hskip1pt(\ref{EqSuperJ}) can be approximately used. The superexchange interaction causes not only the development of AF spin fluctuations but also the stabilization of a singlet GS by the formation of an itinerant singlet or a resonating valence bond (RVB) on each pair of nearest neighbors, as studied in Appendix~\ref{SecRVB}.
This stabilization mechanism is simply the RVB mechanism proposed by Fazekas and Anderson for the Heisenberg model on the triangular lattice.\cite{fazekas}
The stabilization energy is $O(|J|)$ per unit cell; it is also the quenching energy of a spin at a unit cell. Then, 
\begin{align}\label{EqRVBChi}
\left[\tilde{\chi}_s(0)\right]_{T\rightarrow0\hskip1pt{\rm K}}= O(1/|J|).
\end{align}
From Eqs.\hskip1pt(\ref{EqDefTK}) and (\ref{EqRVBChi}), 
$k_{\rm B}T_{\rm K}= O(|J|)$. The fact that $T_{\rm K}>0\hskip2pt$K for any finite $U$ also casts doubt on whether 
the half-filled GS is an insulator when $U$ is finite.
 
In the Heisenberg limit of $U/|t|\rightarrow +\infty$ with $J=-4t^2/U$ kept constant, the half-filled Hubbard model is reduced to the Heisenberg model. 
The GS of the Heisenberg model is a singlet or a doublet according to the number of spins, even or odd, unless $J=0$. This is a proof or at least evidence that the GS degeneracy of the Hubbard model is never infinite even in Heisenberg limit.
This also casts doubt on whether 
the half-filled GS is an insulator in the Heisenberg limit.

It is obvious that  $T_{\rm K}>0$, $\rho(0)>0$, and $\tilde{\Delta}(0)>0$ for the metallic GS while 
$T_{\rm K}=0\hskip2pt$K, $\rho(0)=0$, and $\tilde{\Delta}(0)=0$ for the possible insulating GS.
If an MI transition occurs at $U_c$, the MI transition is inevitably a discontinuous one: Residual entropy jumps at $U_c$. Since Eq.\hskip2pt(\ref{EqRhoDelta}) is satisfied in the metallic GS, one or both of $\rho(0)$ and $\tilde{\Delta}(0)$ also jump at $U_c$.
It is shown in Appendix\hskip2pt\ref{SecRVB} that $U_c/|t|=+\infty$ if the RVB term is only considered as the multisite self-energy beyond the S$^3$A within the KLT; $\rho(0)$ continuously vanishes as $U/|t|\rightarrow +\infty$ while $\tilde{\Delta}(0)$ diverges as $U/|t|\rightarrow +\infty$ and $\tilde{\Delta}(0)=0$ for $U/|t|=+\infty$.

Physical properties of a spin liquid in the Heisenberg model in one dimension can be mapped to those of a TL liquid, i.e., the GS of the Heisenberg model is a TL spin liquid. \cite{TL1,TL2,TL3,TL4} 
This fact implies that the half-filled GS of the Hubbard model is not the Mott insulator but a TL electron liquid for any finite $U$ even in the Heisenberg limit.
When $U/|t| \gtrsim 1$ or $k_{\rm B}T_{\rm K}= O(|J|)$,
the TL electron liquid is a type of RVB electron liquid because it is mainly stabilized by the RVB mechanism.

If no symmetry is broken in two dimensions and higher, e.g., because of frustration and low dimensionality or because of forced constraint of the Hilbert space, it is straightforward to extend  the analyses in this paper to two dimensions and higher.
When $\delta=0^+$, the GS can never be an insulator with a complete gap open even in the Heisenberg limit.
If no symmetry is broken in a spin liquid in the Heisenberg model, 
it is probable that physical properties of the spin liquid in the Heisenberg model can also be mapped to those of an electron liquid in the Hubbard model, as in one dimension, except for conductivity; e.g.,
those of the RVB spin liquid in the Heisenberg model on the triangular lattice \cite{fazekas} must be mapped to those of an RVB electron liquid in the Hubbard model on the triangular lattice.
Any type of spin liquid, in which the third law is not broken and the bosonic spin-excitation spectrum is normal, should be distinguished from the Mott insulator discussed in this paper, in which $T_{\rm K}=0\hskip2pt$K and the third law is broken; $T_{\rm K}=0\hskip2pt$K means or at least implies that there is degeneracy at the zero energy in the bosonic spin excitation.

The Hubbard model in one dimension becomes the Tomonaga-Luttinger (TL) model,
when $E(k)$ given by Eq.\hskip1pt(\ref{EqDispD=1}) is replaced by the dispersion relation composed of two linearized branches defined by
\begin{align}\label{EqDispLinear}
E_{\rm TL}(k) =E(k_{\rm F})
\pm 2|t|a\sin(k_{\rm F}a) (k\mp k_{\rm F}) . 
\end{align}
%
The TL model can be mapped to a boson model; the mapping is rigorous only if the cutoff for Eq.\hskip2pt(\ref{EqDispLinear}) is infinitely large. 
In the boson model, the charge and spin parts are separable from each other, i.e., the charge-spin separation occurs.\cite{Tomonaga,Luttinger,Mattis,comCS-separation}
When $k_{\rm F}=\pi/2a$, a gap opens in the charge excitation because of umklapp processes.\cite{emery} 
The charge gap is as large as 
\begin{align}\label{EqChargeGap}
\Delta_\rho = Ua/(2\pi\alpha),
\end{align}
for a particular $U$, where $1/\alpha$ is the cutoff discussed above; however,
it is not clear whether $\Delta_\rho$ as a function of $U$ is analytic or singular at $U=0$ and whether $\Delta_\rho$ corresponds to $\epsilon_{\rm g}(L)$. 
Even in the gapped phase where the charge gap $\Delta_\rho$ opens, the spin-excitation spectrum is normal: No gap opens in it and no degeneracy exists at the zero energy.
If $T_{\rm K}$ is a well-defined property in the boson model,
this normal spectrum means that $T_{\rm K}>0\hskip2pt$K for the gapped phase. 
In the possible insulating GS of the Hubbard model, on the other hand, $T_{\rm K}=0\hskip2pt$K and, presumably, there exists degeneracy at the zero energy in the bosonic spin-excitation spectrum.
The third law of thermodynamic is not broken in the gapped phase of the boson model, while it is broken in the possible insulating GS of the Hubbard model.
Thus, the nature of the gapped phase of the boson model is different from that of the possible insulating GS of the Hubbard model, which casts doubt on whether the opening of the charge gap $\Delta_\rho$ in the boson model corresponds to the opening of a gap in the fermionic $\rho(\varepsilon)$ of the Hubbard model. 

The argument in the previous paragraph also casts doubt on whether $T_{\rm K}$ is a well-defined property in the boson model and the TL model.
In the Hubbard model, the Kondo effect plays a crucial role in causing any insulating GS to be unstable, at least, when $\delta=0^+$.
There are two possible scenarios for the relevance of the Kondo effect in the TL model:
In a scenario, it is a relevant effect, so that $T_{\rm K}$ and $\tilde{\Sigma}_\sigma(\varepsilon+i0)$ are well-defined properties and no gap opens if $\delta=0^+$, as in the Hubbard model.
In the other scenario, 
it is not a relevant effect, so that neither $T_{\rm K}$ nor $\tilde{\Sigma}_\sigma(\varepsilon+i0)$ is any well-defined property and a gap opens even if $\delta=0^+$; the opening of the charge gap in the TL model does not correspond to the opening of a gap in the fermionic $\rho(\varepsilon)$ of the Hubbard model.
The Hubbard model is a lattice model, while the TL model is essentially a continuous model.
%
Since the Kondo effect, which is a single-site effect, is peculiar to a lattice model, the latter scenario is likelier than the former one; e.g., 
it is shown in Appendix\hskip2pt\ref{SecContinousModel} that
the Kondo effect can play no role in a continuous model.
The equivalence between the Hubbard model and the TL model, in particular, 
as regards the opening of the charge gap should be critically reexamined from a point of view whether or not the Kondo effect is a relevant effect in the TL model.

\section{Conclusion}
\label{SecConclusion}
The half-filled ground state of the Hubbard model is studied by the Kondo-lattice theory within the constrained Hilbert subspace where no symmetry is allowed to be broken. 
The Kondo temperature $T_{\rm K}$ or $k_{\rm B}T_{\rm K}$ is the energy scale of single-site quantum spin fluctuations.
Provided that $T_{\rm K}>0\hskip2pt$K, the single-site self-energy for electrons is normal, so that the ground state is never any insulator with a complete gap open; it is a metal if the multisite self-energy for electrons is continuous and finite at the chemical potential or a zero-gap semiconductor if it is discontinuous or divergent at the chemical potential.
Thus, $T_{\rm K}=0\hskip2pt$K is a necessary condition for that the half-filled ground state is an insulator with a complete gap open.
On the other hand, 
any ground state characterized by $T_{\rm K}=0\hskip2pt$K is infinitely degenerate in the thermodynamic limit, unless the electron filling is empty or completely filled. 
Thus, the only possible insulating half-filled ground state is one such that $T_{\rm K}=0\hskip2pt$K and the third law of thermodynamics is broken in it.

In the presence of an infinitesimal phenomenological perturbation,
definitely $T_{\rm K}>0\hskip2pt$K, unless the onsite $U$ is infinite.
The only possible insulating half-filled ground state, which is characterized by $T_{\rm K}=0\hskip2pt$K, is unstable in the presence of the infinitesimal perturbation, though the prototypic Mott insulator for infinite $U$ is stable.

Though Lieb and Wu's claim that the half-filled ground state in one dimension is the Mott insulator for any nonzero $U$ is based on the rigorous result given by the Bethe ansatz solution, the result is not a proof that the half-filled ground state is an insulator with a complete gap open because the result is not a sufficient condition for it but simply a necessary condition for it.
Since the third law is not broken in the Bethe ansatz solution, which implies that $T_{\rm K}>0\hskip2pt$K even in the absence of the infinitesimal perturbation,
it is doubtful whether the half-filled ground state is the only possible insulating ground state, which is characterized by $T_{\rm K}=0\hskip2pt$K.
It is doubtful whether it is the Mott insulator.

The Hubbard model is a lattice model.
Any single-site property of the Hubbard model can be mapped to its corresponding one in the Anderson model that is self-consistently determined by the Kondo-lattice theory.
When $T_{\rm K}>0\hskip2pt$K, the Kondo effect plays a crucial roles in causing the possible insulating ground state to be unstable. 
Since the Kondo effect, which is a single-site effect, is peculiar to a lattice model, it is unlikely that the Kondo effect is a relevant effect in the Tomonaga-Luttinger model, which is essentially a continuous model.
A gap can only open in the fermionic single-particle excitation spectrum of the Hubbard model if $T_{\rm K}=0\hskip2pt$K and the third law is broken, while a gap opens in the charge excitation spectrum of the Tomonaga-Luttinger model, or the boson model to which the Tomonaga-Luttinger model is mapped, even if the third law is not broken. 
Thus, it is doubtful whether the opening of a gap in the charge excitation spectrum of the Tomonaga-Luttinger model or the boson model is a proof that a complete gap opens in the fermionic single-particle excitation spectrum of the Hubbard model.
It should be reexamined whether the mapping of
the Hubbard model in one dimension to the Tomonaga-Luttinger model and the boson model is really relevant as regards the opening of the gap in the bosonic charge-excitation spectrum.

\section*{Acknowledgments}
The author thanks to T. Hikihara for useful discussions.


\appendix
\section{RVB stabilization effect} 
\label{SecRVB}
In this Appendix,  one dimension, $\left<{\cal N}\right>\hskip-1pt=\hskip-1pt L$, $U/|t|\hskip-1pt \gg\hskip-1pt 1$, and $\delta=0^+$ are assumed and the RVB term is only considered as the multisite self-energy beyond the S$^3$A within the KLT. 
The RVB term is the Fock-type term of the superexchange interaction: 
\begin{align}\label{EqFock0}
\Delta\Sigma_\sigma^{\rm (RVB)}(i\varepsilon_l ,k) &=
\frac{k_{\rm B}T}{L}\sum_{np\sigma^\prime}
\tilde{\phi}_s^2 \frac{1}{4}J_s(k-p)
\bigl({\bm \sigma}^{\sigma\sigma^\prime}
\hskip-3pt\cdot{\bm \sigma}^{\sigma^\prime\hskip-1pt\sigma}\bigr)
\nonumber \\ & \qquad \times
G_{\sigma^\prime}^{\rm (c)}(i\varepsilon_n,p)e^{i\varepsilon_n 0^+},
\end{align}
where 
\begin{align}
J_s (q) = 2J \cos(qa), 
\end{align}
with $J$ given by Eq.\hskip1pt(\ref{EqSuperJ}), 
${\bm \sigma}^{\sigma\sigma^\prime}$ is the $\sigma\sigma^\prime$ component of the Pauli matrix \mbox{${\bm \sigma} =(\sigma_x,\sigma_y,\sigma_z)$}, and $G_{\sigma}^{(c)}(i\varepsilon_n,p)$ is the low-energy coherent part of the total $G_{\sigma}(i\varepsilon_n,p)$, which is given (\ref{EqGreen}).
Since the energy dependence of the superexchange interaction is ignored, the coherent part is only considered in Eq.\hskip1pt(\ref{EqFock0});  
thus, Eq.\hskip2pt(\ref{EqFock0}) is only valid for $|\varepsilon_l |\lesssim k_{\rm B}T_{\rm K}$ if and only if $T_{\rm K}>0\hskip2pt$K.
According to the Ward identity,\cite{ward} $\tilde{\phi}_s$ defined by Eq.\hskip2pt(\ref{EqExpansion}) is the static part of the reducible single-site three-point vertex function in spin channels;
$\tilde{\phi}_s$ is approximately used as its low-energy dynamical part.
Since multisite or intersite effects are perturbatively considered based on the S$^3$A,\cite{Metzner,Muller-H1,Muller-H2,Janis} the two single-site vertex functions should be included in Eq.\hskip1pt(\ref{EqFock0}); there is no double counting in this treatment.

When $\delta=0^+$, the expansion (\ref{EqExpansion}) is relevant.
Then, the coherent part is given by
\begin{align}\label{EqCoherentSimple}
G_{\sigma}^{(c)}(i\varepsilon_l ,k) &=
\frac1{\displaystyle
i \tilde{\phi}_\gamma\varepsilon_l  - E(k) - \Delta\Sigma_\sigma^{\rm (RVB)}(i\varepsilon_l ,k)},
\end{align}
%
and the RVB term is given by
\begin{align}\label{EqFock}
\Delta\Sigma_\sigma^{\rm (RVB)}(i\varepsilon_l ,k) &=
\frac{3}{2} J \tilde{\phi}_\gamma \tilde{W}_s^2 \Xi
\cos(ka).
\end{align}
Here, $\tilde{W}_s$ is the Wilson ratio defined by Eq.\hskip1pt(\ref{EqWilsonRatio}) and
\begin{align}\label{EqXi2}
\Xi &=
\frac1{L}\sum_{k} \cos(ka) f\left[\xi(k)\right],
\end{align}
where $f(\varepsilon)$ is defined by Eq.\hskip2pt(\ref{EqFD-F}) and
\begin{align}\label{EqLowerXi}
\xi(k) &=
- 2 (t^*/\tilde{\phi}_\gamma)\cos(ka),
\end{align}
where
\begin{align}\label{EqTd*}
t^* &= 
t- \frac{3}{4}J\tilde{\phi}_\gamma \tilde{W}_s^2 \Xi.
\end{align}
In the limit of $T\rightarrow0\hskip2pt$K, $\Xi=1/\pi=0.31831\cdots$. 

The coherent part is simply given by
\begin{align}
G_{\sigma}^{(c)}(i\varepsilon_l ,k) &=
\frac1{\tilde{\phi}_\gamma}
\frac1{\displaystyle
i \varepsilon_l  - \xi(k)}.
\end{align}
The bare dispersion relation $E(k)$ is renormalized into $\xi(k)$ by the RVB mechanism.
The DOS of the coherent part is given by
\begin{align}\label{EqFLR0}
\rho^{\rm (c)}(\varepsilon) &=
\frac1{\tilde{\phi}_\gamma L}\sum_{k}
\delta\bigl[\varepsilon- \xi(k)\bigr] ,
\end{align}
which is simply the DOS of the Gutzwiller band;
\begin{align}\label{EqFLR1}
\rho^{\rm (c)}(\varepsilon) &=
(\alpha_\rho/|t^*|)
\left[ 1 + O\bigl(\varepsilon^2\bigr)\right], 
\end{align}
with $\alpha_\rho=O(1)$, for $|\varepsilon|\lesssim |t^*|/\tilde{\phi}_\gamma$.
The bandwidth of $\xi(k)$ or $\rho^{\rm (c)}(\varepsilon)$ is 
\begin{align}\label{EqEffectiveBandwidth}
W^*= 2|t^*|/\tilde{\phi}_\gamma,
\end{align}
which corresponds to $k_{\rm B}T_{\rm K}$.
It should be noted that
\begin{align}
W^* \rightarrow 3|J|\tilde{W}_s^2\Xi/4,
\end{align}
if $\tilde{\phi}_\gamma\rightarrow +\infty$, i.e., $W^*$ is nonzero even if $\tilde{\phi}_\gamma\rightarrow +\infty$, while $\rho^{\rm (c)}(\varepsilon)|t| \rightarrow 0$ if $\tilde{\phi}_\gamma\rightarrow +\infty$.
It is interesting that, unless $J=0$, the bandwidth $W^*$ of the Gutzwiller band is nonzero even though the band itself is vanishing.

The expansion coefficients $\tilde{\phi}_\gamma$ and $\tilde{\phi}_s$, which are single-site properties, should be self-consistently calculated with the RVB term.
In principle, this problem should be solved by self-consistently determining and solving the AM, as discussed in \mbox{Sec.\hskip2pt\ref{SecKL-theory}}. However, we take a qualitative approach to this problem.
%
If the mapping condition is satisfied, the probabilities of empty or double occupancy are the same as each other between the Hubbard model and the mapped AM.
When the virtual processes allowing empty and double occupancies, from which the RVB mechanism arises, are considered, 
the probability is $O\bigl[t^2/U^2\bigr]$ in the Hubbard model.
On the other hand, we assume on the basis of \mbox{Eqs.\hskip1pt(\ref{EqRhoDelta})}, (\ref{EqFLR1}), and (\ref{EqEffectiveBandwidth}) that
\begin{align}\label{EqFLR2}
\tilde{\Delta}(\varepsilon) &=
\left\{\begin{array}{cc}
\displaystyle
|t^*|/(\pi\alpha_\rho), 
& \displaystyle
|\varepsilon| \le W^*/2 
\vspace{2pt}\\
0, 
& \displaystyle
|\varepsilon| > W^*/2
\end{array} \right. ,
\end{align}
for the mapped AM.
The probability is $O\bigl[\tilde{\Delta}(0)/U^2\bigr]\times O\bigl(W^*\bigr)$ in the AM.
The two probabilities should be equal to each other:
\begin{align}\label{EqFLR3}
O\bigl[t^2/U^2\bigr] &=
O\bigl[\tilde{\Delta}(0)W^*/U^2\bigr].
\end{align}
When $U/|t|\gg 1$ and $1<\tilde{W}_s<2$ are assumed,
it follows from \mbox{Eqs.\hskip1pt(\ref{EqTd*})}, (\ref{EqFLR2}), and (\ref{EqFLR3}) that
\begin{align}\label{EqFLR4}
\tilde{\phi}_\gamma =O\bigl(U/|J|\bigr) ,
\end{align}
it follows from Eq.\hskip2pt(\ref{EqTd*}) and (\ref{EqFLR4}) that
\begin{align}
|t^*| = O\bigl(U\bigr),
\end{align}
and it follows from Eqs.\hskip2pt(\ref{EqRhoDelta}) and (\ref{EqFLR1}) that
\begin{align}\label{EqFLR6}
\rho^{\rm (c)}(0)=1/\bigl[\pi\tilde{\Delta}(0)\bigr] = O\bigl(1/U\bigr). 
\end{align}
Then,
\begin{align}\label{EqRhoUInf}
\lim_{U/|t|\rightarrow+\infty}\rho(0) =0,
\end{align}
and
\begin{align}\label{EqDeltaUInf}
\lim_{U/|t|\rightarrow+\infty}\tilde{\Delta}(0)/|t| =+\infty.
\end{align}
An MI transition occurs at $U_c/|t|=+\infty$ as a function of $U/|t|$. The MI transition is a discontinuous one, as discussed in Sec.\hskip2pt\ref{SecDiscussion}.

Since  
%
$W^*=2|t^*|/\tilde{\phi}_\gamma = O(|J|)$, 
%
$T_{\rm K}$ is estimated to be
\begin{align}\label{EqTK=J}
k_{\rm B}T_{\rm K} = O(|J|).
\end{align}
%
%
%
If the RVB term is once self-consistently considered, even if $\delta=0$ is assumed, Eqs.\hskip2pt(\ref{EqDeltaUInf}) and (\ref{EqTK=J}) guarantee that $T_{\rm K}$ is nonzero even for $U/|t|\gg 1$, though $T_{\rm K}\rightarrow 0\hskip2pt$K as $U/|t|\rightarrow+\infty$.

According to the analysis above, it is plausible that, in general,  the GS can never be an insulator provided that the virtual processes allowing empty and double occupancies are possible, i.e., provided that $U$ is finite.

It is straightforward to show that anomalous terms proportional to
\begin{align}\label{EqAnomalDelta}
\varepsilon\left[\ln(\varepsilon+i0)+
\ln(-\varepsilon-i0) \right] =
2\varepsilon\ln|\varepsilon| - i\pi|\varepsilon|,
\end{align}
appear in the multisite self-energy higher order in $J_s(q)$. 
Since the anomalous terms are continuous at $\varepsilon=0$, they can cause no gap to open in $\rho^{\rm (c)}(\varepsilon)$ or $\rho(\varepsilon)$.
Within a preliminary study, any anomalous term can be found such that it causes a gap or a zero-gap to open.
Since Eq.\hskip2pt(\ref{EqAnomalDelta}) is anomalous at $\varepsilon=0$, the half-filled GS in one dimension is never a normal FL.
If no gap really opens as discussed in Sec.\hskip2pt\ref{SecDiscussion}, it is an RVB type of TL liquid.

\section{Irrelevance of the Kondo effect in a continuous model}
\label{SecContinousModel}
We consider a continuous model in one dimension:
\begin{align}\label{EqContM1}
{\cal H}_a &=
\hbar v_{\rm F} \sum_\sigma
\int_{0}^{x_1} \hskip-5pt dx \hskip2pt \psi_\sigma^\dag (x)
\hskip-1pt\left(-i\frac{\partial \phantom{x}}{\partial x}- k_{\rm F}\right)\hskip-1pt
\psi_\sigma (x),
\nonumber \\ &\quad
+ \frac1{2}g_1\sum_\sigma
\int_{0}^{x_1} \hskip-5pt dx \hskip2pt
\psi_\sigma^\dag (x)\psi_{-\sigma}^\dag (x)
\psi_{-\sigma} (x) \psi_\sigma (x),
\end{align}
where $\psi_\sigma^\dag (x)$ and $\psi_\sigma(x)$ are fermionic field operators.
Since we are interested in the correspondence of this model to the Hubbard model, $\hbar v_{\rm F} = 2|t|a$, $x_1=La$, $k_{\rm F}=\pi/a$, and $g_1=Ua$ are assumed. 
When the periodic boundary condition is assumed and 
\begin{align}
\psi_\sigma (x)=
\frac1{\sqrt{La}} \sum_{k} e^{ikx} a_{k\sigma}, 
\end{align}
is used, it follows that 
\begin{align}\label{EqContM2}
{\cal H}_a &=
\sum_{k\sigma} \hbar v_{\rm F} (k -k_{\rm F}) a_{k\sigma}^\dag a_{k\sigma}
\nonumber \\ & \quad
+ \frac{g_1}{2La}\sum_{kpq\sigma}a_{k+q\sigma}^\dag a_{p-q-\sigma}^\dag a_{p-\sigma} a_{k\sigma}.
\end{align}

When the continuous model (\ref{EqContM1}) or (\ref{EqContM2}) is extended into a two-band continuous model in which the dispersion relation is given by Eq.\hskip2pt(\ref{EqDispLinear}), it is the Tomonaga-Luttinger (TL) model.
When $U$ is nonzero in the Hubbard model, electrons and holes with energy $|\varepsilon-\mu|\lesssim U$ are virtually excited in the GS; 
$\left<n_{k\sigma}\right> =O(1)$ for $k$ such that $0<E(k)-\mu\lesssim U$,
and $\left<1-n_{k\sigma}\right> =O(1)$ for $k$ such that $-U \lesssim E(k)-\mu<0$, where $n_{k\sigma}$
%
%
is the number operator for the Hubbard model in the wave-number representation.
On the other hand,
Eq.\hskip2pt(\ref{EqDispLinear}) can only accurately describe electrons in the vicinity of the Fermi level, i.e., electrons with $|k- k_{\rm F}|a\ll 1$ or $|k+ k_{\rm F}|a\ll 1$.
Thus, the correspondence is only accurate in the weak-coupling case of $U/|t|=2g/(\hbar v_{\rm F}) \ll 1$.
It is questionable how accurately and relevantly electron correlation in the strong-coupling Hubbard model with $U/|t|\gg 1$ can be treated by the strong-coupling TL model with $g/(\hbar v_{\rm F}) \gg 1/2$. When $U/|t|\gtrsim 2$ in the Hubbard model, for example, either of the splitting into the upper and lower Hubbard bands, the superexchange interaction, and the RVB mechanism is a relevant effect; however, neither of them can be treated by the TL model with $g/(\hbar v_{\rm F}) \gtrsim 1$.

In general, 
when the dispersion relation is linear in $k$ in a model such as one defined by Eq.\hskip2pt(\ref{EqContM2}), the model is essentially a continuous model.
The TL model is essentially a continuous model,
even if umklapp processes are considered as one of lattice effects.


We denote the Green function and the self-energy in the continuous model (\ref{EqContM1}) or (\ref{EqContM2}) by
$G_\sigma(i\varepsilon_l , x-x^\prime)$ 
and
$\Sigma_\sigma(i\varepsilon_l , x-x^\prime)$,
respectively, in the real-space representation.
When we follow the definition of the single-site diagram for the Hubbard model,
the {\it single-site} or local diagram should be defined as a diagram such that
it includes only $g$-lines and local lines of the Green function, which is simply $G_\sigma(i\varepsilon_l , x-x^\prime=0)$.
If at least a {\it multisite} or non-local line of $G_\sigma(i\varepsilon_l , x- x^\prime\ne 0)$
is included in a diagram, the diagram is a non-local one.
Thus, the self-energy can also be decomposed into the local $\tilde{\Sigma}_\sigma(i\varepsilon_l )$ and the non-local $\Delta\Sigma_\sigma(i\varepsilon_l ,x-x^\prime)$ in a formal way:
\begin{align}
\Sigma_\sigma(i\varepsilon_l , x \hskip-1pt-\hskip-1pt x^\prime) &= 
\tilde{\Sigma}_\sigma(i\varepsilon_l )\delta_{x \hskip-1pt-\hskip-1pt x^\prime}
+ \Delta\Sigma_\sigma(i\varepsilon_l ,x \hskip-1pt-\hskip-1pt x^\prime),
\end{align}
where $\delta_{x}$ is a Kronecker-like delta defined for real $x$ by
\begin{align}
\delta_{x} &=
\left\{\begin{array}{cc}
1, & x=0 \\
0, & x\ne 0
\end{array}\right. .
\end{align}
It should be noted that $\delta_{x}$ is never the delta function.
Since $x$ is a continuous variable, $\delta_{x}=0$ in a practical or physical sense. 
It is obvious that 
the local $\tilde{\Sigma}_\sigma(i\varepsilon_l )\delta_{x-x^\prime}$ can play no role in the continuous model. 
The Kondo effect can play no role in any continuous model.

The TL model is essentially a two-band continuous model.
Thus, the proof of this paper for the Hubbard model that no complete gap can open because of the Kondo effect in the presence of $\delta=0^+$ never contradicts the opening of the charge gap in the TL model or the boson model, where the Kondo effect is irrelevant.

\section{Instability of the Mott insulator in infinite dimensions}
\label{SecInfDim}
%
When the Hubbard model in $D$ dimensions is studied, e.g.,
$t/\sqrt{D}$ is substituted for $t$ in Eq.\hskip1pt(\ref{EqHubbardH}).
Then 
\begin{align}
E_D({\bf k}) = - \frac{2t}{\sqrt{D}}\sum_{\nu=1}^{D}
\cos(k_\nu a),
\end{align}
is substituted for Eq.\hskip1pt(\ref{EqDispD=1}). 
The effective bandwidth of $E_D({\bf k})$ is $O(|t|)$ for any $D$.
The single-site $\tilde{\Sigma}_\sigma(i\varepsilon_l)$ is of the leading order in $1/D$, the multisite $\Delta\Sigma_\sigma(i\varepsilon_l, {\bf k})$ is of higher order in $1/D$, and the conventional Weiss mean field (MF), which is a multisite effect, can be of the leading order in $1/D$.\cite{comLeadingMF}
If the single-site term is rigorously considered and 
no multisite term is considered in a theory, the theory is within the S$^3$A.\cite{Metzner,Muller-H1,Muller-H2,Janis}
The KLT is in a sense $1/D$ expansion theory based on the S$^3$A to include multisite terms.

First we consider the half-filled GS in the S$^3$A.
Since no conventional Weiss MF is considered, no symmetry can be broken in the S$^3$A. Then, the whole analysis in Sec.\hskip2pt\ref{SecResult} is also valid within the S$^3$A.
If $\delta=0^+$ is assumed, $T_{\rm K}>0\hskip1pt$K and the GS is a normal FL, though $T_{\rm K}$ may be infinitesimally low as discussed below.
According to Eqs.\hskip1pt(\ref{EqDOS1}), (\ref{EqMuEdU}), and (\ref{EqNatureSA}), 
\begin{align}\label{EqS3ARho}
\rho(0) = \frac1{L}\sum_{\bf k} \delta\left[-E_D({\bf k})\right]>0.
\end{align}
Thus, $\rho(0)$ does not depend on $U$.\cite{comRhoConst}
Because of Eq.\hskip1pt(\ref{EqRhoDelta}), 
%
$\tilde{\Delta}(0)=1/[\pi\rho(0)]>0$
%
does not depend on $U$ either.
The self-energy can be expanded as in Eq.\hskip2pt(\ref{EqExpansion}).
According to the FL theory,\cite{luttinger} the specific-heat coefficient and the static homogeneous spin susceptibility are given by
\begin{align}
\gamma = (2/3) \pi^2 k_{\rm B}^2 \tilde{\phi}_\gamma\rho(0),
\end{align}
and 
\begin{align}
\bigl[\chi_s(0,{\bf q})\bigr]_{|{\bf q}|\rightarrow 0}
= 2 \tilde{\phi}_s\rho(0),
\end{align}
respectively.
A discontinuous MI transition occurs at $U_c/|t|=+\infty$ between a normal FL and the prototypic Mott insulator in case of $\delta=0^+$. 

According to theories based on the DMFT,\cite{kotliar,moeller,bulla}
on the other hand, a discontinuous MI transition occurs at finite $U$ in case of $\delta=0$.
When $U$ increases, a complete gap opens at $U_{c2}=O(|t|)$; both of $\gamma$ and $\bigl[\chi_s(0,{\bf q})\bigr]_{|{\bf q}|\rightarrow 0}$ diverge as $U\rightarrow U_{c2}-0$.
When $U$ decreases, the complete gap closes at $U_{c1}=O(|t|)$, which is smaller than $U_{c2}$.


In the DMFT, all the single-site terms are rigorously considered as the {\it dynamical} MF, and any conventional Weiss MF, which is a {\it static} MF, is not considered. 
Thus, the DMFT and the KLT are exactly equivalent to each other within the S$^3$A.
According to the study based on the KLT in this paper,
the discontinuous MI transition characterized by two $U_c$'s, $U_{c1}$ and $U_{c2}$, must be one between a normal FL, which is characterized by $T_{\rm K}>0\hskip2pt$K, and the Mott insulator, which is characterized by $T_{\rm K}=0\hskip2pt$K.
The fact that no order parameter can appear in the DMFT and both of $\gamma$ and $\left[\chi_s(0,{\bf q})\right]_{|{\bf q}|\rightarrow 0}$ diverge as $U\rightarrow U_{c2}-0$ is a proof or at least a strong piece of evidence that $T_{\rm K}=0\hskip2pt$K and residual entropy remains in the insulting GS or the Mott insulator in the DMFT.

According to the study in Sec.\hskip2pt\ref{SecResult} of this paper, 
$T_{\rm K}>0\hskip2pt$K for $\delta>0$.
Then, the following two cases are possible for the dependence of $T_{\rm K}$ on $\delta$:
\begin{subequations}
\begin{align}\label{EqTK-P}
\lim_{\delta\rightarrow 0}T_{\rm K} >0\hskip2pt{\rm K},
\end{align}
and
\begin{align}\label{EqTK-Z}
\lim_{\delta\rightarrow 0}T_{\rm K} =0\hskip2pt{\rm K}.
\end{align}
\end{subequations}
Based on the result of the DMFT for $\delta=0$, 
it is anticipated that Eq.\hskip1pt(\ref{EqTK-P}) is satisfied for $U<U_{c2}$ while Eq.\hskip1pt(\ref{EqTK-Z}) is satisfied for $U\ge U_{c2}$.
It is desirable to confirm this dependence of $T_{\rm K}$ on $\delta$ and $U$ by a theory based on the DMFT.
If $\delta=0^+$ is assumed in the theory, the GS of the theory is not the Mott insulator but a normal FL.

Since a normal FL with $T_{\rm K}=+0\hskip1pt$K behaves like the Mott insulator in a high-$T$ phase at $T\gg T_{\rm K}=+0\hskip1pt$K,
there is no difference between the normal FL with $T_{\rm K}=+0\hskip2pt$K and the Mott insulator, excepting at $T=0\hskip1pt$K. 
There is no practical difference between physical properties of the phase with $T_{\rm K}=+0\hskip1pt$K and those of the phase with $T_{\rm K}=0\hskip1pt$K. There is no significant inconsistency between the theoretical result based on the DMFT\cite{kotliar,moeller,bulla} and the analysis based on the KLT in this paper.

Next we consider the half-filled GS beyond the S$^3$A; but only the RVB term is considered beyond it within the KLT, as in Appendix\hskip2pt\ref{SecRVB}.
When $U/|t| \gg 1$, the superexchange-interaction constant is given by 
\begin{align}
J_D = -4t^2/(DU), 
\end{align}
between nearest neighbors. 
The stabilization energy by the RVB mechanism is
$O(|J_D|)$ per unit cell, so that
\begin{align}\label{EqTKRVB}
k_{\rm B}T_{\rm K} =O(|J_D|)= O[t^2/(DU)].
\end{align}
The RVB mechanism is of higher order in $1/D$.
Following Appendix\hskip2pt\ref{SecRVB}, it is straightforward to show that $\rho(0)|t| \rightarrow 0$ as $U/|t|\rightarrow +\infty$ for any finite $D$; $\rho(0)$ is not given by Eq.\hskip1pt(\ref{EqS3ARho}). 
If $\rho(0)|t|\ll 1$ for finite $D$ and $U/|t|\gtrsim 1$, the RVB mechanism is crucial, so that the GS is simply an RVB type of normal FL.
On the other hand, if the limit of $D\rightarrow +\infty$ is followed by that of $U/|t|\rightarrow +\infty$, the RVB term vanishes, so that
$\rho(0)$ is given by Eq.\hskip1pt(\ref{EqS3ARho}).
The constancy of $\rho(0)$ as a function of $U$ is not an property of $D\rightarrow +\infty$ dimensions but one in the S$^3$A.

Thirdly we consider the true half-filled GS.
When $U/|t|\gg 1$, the N\'{e}el temperature $T_{\rm N}$ is given by
\begin{align}
k_{\rm B}T_{\rm N} = D|J_D|/2= 2t^2/U,
\end{align}
in the MF approximation, which is rigorous in the limit of $D\rightarrow+\infty$. 
The Weiss MF of magnetism is a leading-order effect in $1/D$.\cite{comLeadingMF}
Since 
\begin{align}
T_{\rm N} \gg T_{\rm K}= O[t^2/(DU)], 
\end{align}
the GS is a local-moment type of AF magnet. 

In a high-$T$ phase at $T\gg T_{\rm K}$, the susceptibility of the mapped AM obeys the Curie-Weiss (CW) law:
\begin{align}\label{EqAM-CW}
\tilde{\chi}_s(0)\simeq 1/[k_{\rm B}(T+T_{\rm K})]. 
\end{align}
According to the KLT,
the effective Fermi energy $E_{\rm F}^*$ discussed in Sec.\hskip2pt\ref{SecIntroduction} is as large as 
%
$E_{\rm F}^*\simeq k_{\rm B}T_{\rm K}$.
%
In a high-$T$ phase at $T>T_{\rm N}$ and $T\gg E_{\rm F}^*/k_{\rm B}$ or $T\gg T_{\rm K}$, 
\begin{align}\label{EqSusHubA}
\chi_s(0,{\bf q}) &=
\tilde{\chi}_s(0)/\bigl[1 - (1/4)J_s({\bf q})\tilde{\chi}_s(0) \bigr]
\nonumber \vspace{5pt}\\ &\simeq
(1/k_{\rm B})/\bigl[T + T_{\rm K} - J_s({\bf q})/(4k_{\rm B})\bigr],
\end{align}
where
\begin{align}
J_s({\bf q}) = 
2J_D \sum_{\nu=1}^{D}\cos(q_\nu a). 
\end{align}
%
The susceptibility obeys the CW law for any ${\bf q}$ because of the $T$ dependence of $\tilde{\chi}_s(0)$; the Weiss constant depends on ${\bf q}$.
Since electrons behave like local moments, the high-$T$ phase can be regarded as the Mott insulator. When $U/|t|\gg 1$, the GS is an AF type of Mott insulator.

When $U/|t|\ll 1$, the GS is also an AF insulator.
Since 
\begin{align}
T_{\rm N} \ll T_{\rm K}=O(|t|/k_{\rm B}), 
\end{align}
it is an itinerant-electron type of AF magnet.
The spin susceptibility obeys the CW law at $T_{\rm N}<T\ll T_{\rm K}$ for particular ${\bf q}$ such as ${\bf q}\simeq(\pi/a)(\pm 1, \pm 1, \cdots, \pm 1)$ because of the nesting of the FS.
When $U/|t|\ll 1$, the GS is not any type of Mott insulator.

Lastly we consider the half-filled or non-half filled GS for $U/|t|\gtrsim 
1$ and finite $D$ within a treatment where
the single-site $\tilde{\Sigma}_\sigma(i\varepsilon_l)$ is rigorously considered, the RVB term is considered, all the other multisite terms are not considered, and no Weiss MF is considered.
Within this treatment,
the GS is an RVB type of normal FL. 
This type of normal FL is so useful that it can be used as a {\it unperturbed} state to study high-$T_c$ superconductivity in the vicinity of an AF type of Mott insulator.\cite{FJO-SC2}



\begin{thebibliography}{9}
\bibitem{comInfU-1D}
When $U$ is infinite and the transfer integral is nonzero only between nearest neighbors in the one-dimensional Hubbard model, and if no electron reservoir exists, the charge-spin separation is complete in a sense such that the wave function of any eigenstate is a direct product of that of $N$ free spinless fermions and that of $N$ free spins. Wen $N\ne L$, the GS is an exotic metal where residual entropy is $k_{\rm B}\ln 2$ per electron and all the bosonic spin excitations are degenerate at the zero energy, though the bosonic charge-excitation spectrum is normal; 
however, no charge-spin separation occurs in the fermionic single-particle excitation.
\bibitem{comMS}
Rigorously speaking, 
$\epsilon_{\rm g}(N)=0$ is a sufficient condition for the GS being a non-insulator, i.e., a metal, a semi-metal, or a zero-gap semiconductor.
\bibitem{comBandInsulator}
Quantum fluctuations of the total electron number vanish in a band insulator. Thus, the band insulator is rigid, if and only if each and every band is empty or completely filled. It becomes a metal when an infinitesimal number of electron or holes are doped.
\bibitem{comExoticMetal}
If $T_{\rm K}$ is defined for the exotic metal discussed in the comment,\cite{comInfU-1D} $T_{\rm K}=0\hskip2pt$K because of the degeneracy at the zero energy in the bosonic spin excitation.
\bibitem{comCS-separation}
The charge-spin separation in the boson model and the TL model occurs in a different way from that in the Hubbard model with $U/|t|=+\infty$, which is discussed in the comment.\cite{comInfU-1D}
\bibitem{comLeadingMF}
Three types of Weiss MF are of the leading order in $1/d$: CDW and BCS or isotropic $s$-wave superconductivity for $U/|t|<0$, and magnetism for $U/|t|>0$.
\bibitem{comRhoConst}
In the S$^3$A, $\rho(0)$ does not depend on $U$ for any filling.
\end{thebibliography}
\end{document}